# An FPGA Based Fast Linear Discharge Method for Nuclear Pulse Digitization

Xiaoguang Kong, Yonggang Wang, Liwei Wang, Yong Xiao, Jie Kuang

*Abstract*—Inspired by Wilkinson ADC method, we implement a fast linear discharge method based on FPGA to digitize nuclear pulse signal. In this scheme, we use a constant current source to discharge the charge on capacitor which is integrated by the input current pulse so as to convert the amplitude of the input nuclear pulse into time width linearly. Thanks to the high precision of TDC measurement that we have achieved in FPGA, we can increase the current value of the discharge to make the discharge time short, so as to obtain a small measurement of dead time. We have realized a single channel fast linear discharge circuit which contains only one dual supply amplifier, two resistors and one capacitor. The rest part can be implemented in an FPGA (Field Programmable Gate Array). Leakage current from the sensor would cause the base line drifting slowly, which can influence the measuring precision. Our method solves this problem without losing the linearity of measurement. We have built the circuit and experimental setup for evaluation. Using them to measure energy spectrums of PET detectors of PMT coupled with LYSO and LaBr3 crystal, the energy resolution is 12.67% and 5.17% respectively. The test results show that our circuit is rather simple, stable and conducive for multi-channel integration.

## I. INTRODUCTION

PARTICLE detectors often require front-end electronics to digitize the output analog pulse signals with short dead time, high energy resolution, large measurable dynamic range and capability for multi-channel integration. For example, high resolution PET detector module based on large continuous crystals need such a front-end electronics [1]. Traditional methods, such as waveform digitization scheme include high speed ADC sampling, Time over threshold are not well suitable for the case. On the basis of Wilkinson ADC scheme, combining our high precision time-to-digital (TDC) technique, a field programmable gate array (FPGA) based fast linear discharge method is proposed in this paper. Its advantages of short dead time, high precision and simplicity makes it very suitable for the PET detectors

Manuscript received January 30, 2018. This work was supported by National Natural Science Foundation of China (NSFC) under Grants 11475168 and 11735013.

Authors are with Department of Modern Physics, University of Science and Technology of China, Hefei, Anhui 230026, China. Corresponding author: wangyg@ustc.edu.cn.

## II. METHOD AND EXPERIMENTAL CIRCUIT

The principle of Wilkinson ADC is first to integrate the charge of pulses on a capacitor, which final voltage is proportional to the charge of the input signal. After the charge completion, a current source linearly discharge the charge on capacitor until the voltage on capacitor linearly reducing to zero. The discharge time has a linear relation with the total charge on the capacitor. By measuring the discharge time using a high-speed counter, the nuclear charge is measured. Based on this idea, we integrate the input pulse on a capacitor with an inverting amplifier, nearly at the same time, a constant current source starts to discharge the charge linearly. The discharge time is still linear to the charge of input pulse as long as it is 3 times greater than the decay time of input pulse. By conversion, the charge of the input signal can be obtained by measuring the discharge time of the signal. Due to leakage current and electronic noise, the charge on the capacitor will accumulate without a feedback resistor when signal has not arrived, which will lead to an obvious baseline shifting at the arrival of event and affect the measurement precision. If we use a parallel resistor on capacitor to drain the charge to stabilize the baseline, then discharge current won't be constant when measuring the signal, which will cause non-linear error between discharge time and pulse amplitude. In order to solve this problem, we use a tri-state gate of FPGA to stabilize the baseline without affecting the linearity of the measurement. Subsequent experiments show that the circuit is not only simple in structure, but also flexible and reliable.

We have realized the linear discharge circuit of single channel shown in Fig. 1. The circuit is composed of a dual

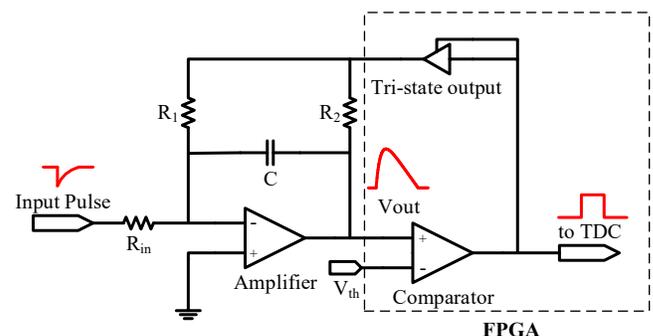

Fig. 1. The basic circuit of fast linearly discharge method. The comparator and TDC module are implemented in the FPGA.

supply amplifier, two feedback resistors, one feedback capacitor and an FPGA. An LVDS comparator and a TDC

block are implemented in the FPGA. The output of amplifier is connected to the positive input port of LVDS comparator directly to compare with a fixed low threshold voltage. The output of the comparator is connected to the middle of two feedback resistors by a tri-state output buffer of FPGA.

When a signal does not arrive, the charge accumulated on capacitor by noise or leakage current will be released through the feedback resistors $R_1$ and $R_2$, which keeps the voltage on capacitor stable below the threshold voltage. And the tri-state output of FPGA is high resistance. When a signal arrives, the voltage accumulates on capacitor beyond the threshold voltage, causing the comparator to flip and the output of the tri-state output to convert from high resistance to digital high output. As a result, the middle of two feedback resistors become pulled up rather than hanging. As we know, FPGA digital high output supply a fixed voltage and the input negative port of amplifier is virtual short to ground, so we can get a constant voltage difference on resistor $R_1$, so the current on resistor $R_1$ is constant which will discharge the charge on capacitor linearly. Meanwhile, $R_2$ works as a pull-up resistor and does not drain the charge on capacitor at all. The discharge process come to an end when the voltage on capacitor cross the threshold voltage again. Then the LVDS comparator flips which leading to the tri-state output return to high resistance. All in all, an input pulse from sensor will lead to a rectangular pulse whose width is linear to the charge of input pulse. Measuring the rectangular pulse width using a TDC in FPGA will obtain the amplitude of input pulse we want. Fig. 2 shows an example of test process.

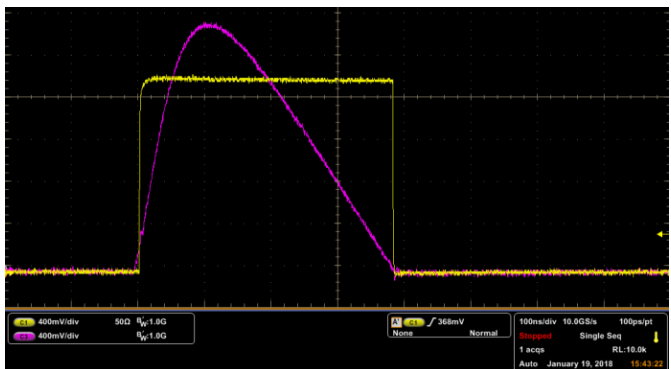

Fig. 2. An oscilloscope scope capture of fast linearly discharge circuit with voltage on capacitor (magenta) and output of LVDS comparator (yellow).

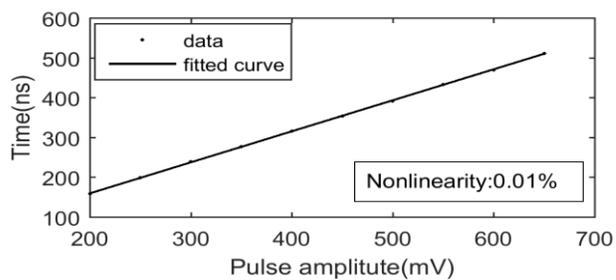

(a)

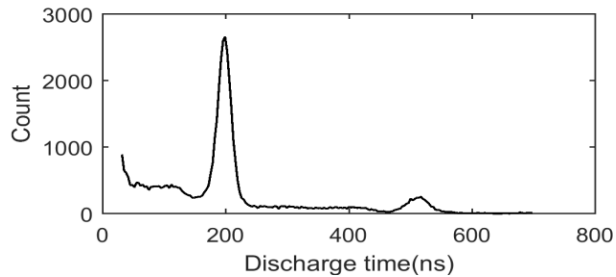

(b)

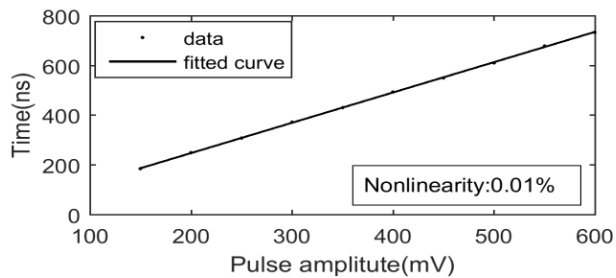

(c)

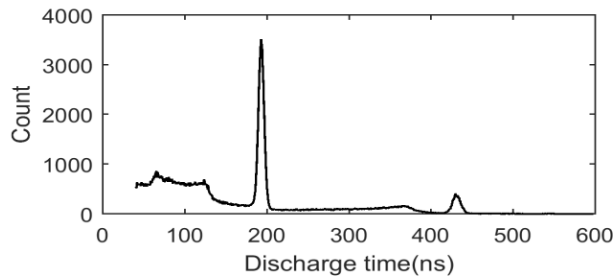

(d)

Fig. 3. (a) Linearity of test circuit ($R_{in}$=100Ω, $R_1$=2.4kΩ, $R_2$=1 kΩ, C=80pF). (b) Measured energy spectrum of $^{22}$Na with LYSO crystal. (c) Linearity of test circuit ($R_{in}$=100 Ω, $R_1$=3.9 kΩ, $R_2$=1 kΩ, C=80pF). (d) Measured energy spectrum of $^{22}$Na with LaBr$_3$ crystal.

## III. Test Result

A realistic board with Xilinx Kintex-7 FPGA was built to evaluate the performance of fast linear discharge method. The detector we use is HAMAMATSU R9800 PMT together with large continuous crystals LYSO ($12mm \times 12mm \times 10mm$) and LaBr$_3$ ($\Phi 22mm \times 10mm$). A dual supply inverting amplifier AD8065, one capacitor and two resistors make up the analog module outside the FPGA. The energy spectrum of $^{22}$Na γ source is measured with this circuit.

Test result (Fig. 3) shows that the nonlinearity of our circuit can be reduced to 0.01%. The 511KeV and 1274KeV energy peak is visible so we can calibrate the conversion between energy and discharge time with them. After calculation, we can get energy resolutions of 511Kev peak 12.67% of LYSO crystal and 5.17% with LaBr$_3$. The test performance evident that this circuit works well as expected.

## IV. Conclusion

We have realized a single channel fast linear discharge method to digitize nuclear pulse with a simple circuit structure combined with an FPGA. The amplitude of nuclear pulse from detector is converted into pulse width for subsequent measurement and processing. The test result shows that this scheme has advantages of high linearity and high energy resolution. Compared with traditional amplitude digitization scheme, this method has a shorter dead time and a larger measurable dynamic range, and consist of fewer passive components which is conducive to multi-channel and high event rates applications.

REFERENCES

[1] Wang Yonggang, Zhu Wensong, and Chen Jun, "A novel nuclear pulse digitizing scheme using time over dynamic threshold", 2011 IEEE Nuclear Science Symposium Conference Record, N50-5, p2174-2179.